\documentclass[%
 aip,
 amsmath,amssymb,
 reprint,%
]{revtex4-1}

\usepackage{graphicx}
\usepackage{dcolumn}
\usepackage{bm}
\usepackage{titlesec}

\usepackage{tablefootnote}
\usepackage[utf8]{inputenc}
\usepackage[T1]{fontenc}
\usepackage{mathptmx}
\usepackage{etoolbox}
\usepackage{xspace}
\usepackage{color,soul}
\usepackage{textcomp}
\usepackage[export]{adjustbox}
\usepackage{setspace}

\newcommand{\HCNH}{HCNH$^{+}$\xspace}

\makeatletter
\def\@email#1#2{%
 \endgroup
 \patchcmd{\titleblock@produce}
  {\frontmatter@RRAPformat}
  {\frontmatter@RRAPformat{\produce@RRAP{*#1\href{mailto:#2}{#2}}}\frontmatter@RRAPformat}
  {}{}
}%
\makeatother
\begin{document}

\preprint{AIP/123-QED}

\title{Hyperfine-Resolved Rotational Spectroscopy of \HCNH}
\author{Weslley G. D. P. Silva}
\author{Luis Bonah}
\author{Philipp C. Schmid}%
\author{Stephan Schlemmer}%
\author{Oskar Asvany*}
 \email{asvany@ph1.uni-koeln.de}
\affiliation{ 
I. Physikalisches Institut, Universit\"at zu K\"oln Z\"ulpicher Str. 77, 50937 K\"oln, Germany
}%

\date{\today}
             
\begin{abstract}
    The  rotational spectrum of the molecular ion \HCNH is  revisited  using  double-resonance spectroscopy in an ion trap apparatus,
    with six transitions  measured between 74 and 445~GHz.
    Due to the cryogenic temperature of the trap,
    the hyperfine splittings caused by the $^{14}$N quadrupolar nucleus were resolved for transitions up to $J=4 \leftarrow 3$,
    allowing for a refinement of the spectroscopic parameters previously reported, especially the quadrupole coupling constant $eQq$.
\end{abstract}

\maketitle

Protonated hydrogen cyanide (\HCNH) is a linear, closed-shell molecular ion, 
which plays an important role in the chemistry of the interstellar medium (ISM), 
being the main precursor for the formation of neutral HCN and HNC \cite{her78}. 
\HCNH has been extensively studied in both laboratory and space. In the laboratory, 
\HCNH was investigated by rotationally-resolved infrared spectroscopy\cite{alt84}, 
followed by pure rotational spectroscopic studies spanning from the microwave\cite{aka98} to the 
sub-millimeterwave\cite{bog85,ama06} spectral region. In space, \HCNH has been detected across 
several interstellar regions based on its pure rotational fingerprints \cite{ziu86,ziu92,que17}. 
In the observations toward the Taurus molecular cloud (TMC-1),  a dense and cold region in the ISM, 
Ziurys et al.\cite{ziu92} observed the three $^{14}$N quadrupolar hyperfine components of the $J= 1 \rightarrow 0$ 
transition around 74 GHz for the first time. A value for the quadrupole coupling 
constant $eQq$= -0.49(7) MHz was derived from these observations. 
Up to date, 
no hyperfine splittings could be resolved for any transitions of \HCNH in the laboratory.

In this communication, we report the first hyperfine-resolved rotational spectrum of \HCNH measured in the laboratory. 
Six transitions were recorded between 74 and 445~GHz using double-resonance spectroscopy. 
Hyperfine splittings due to the $^{14}$N quadrupolar nucleus were resolved for transitions up to $J=4 \leftarrow 3$. 
The measurements presented here were carried out using the 4~K cryogenic ion trap 
instrument called COLTRAP~\citep{asv10,asv14}. The \HCNH ions were created inside a storage ion source via electron impact 
ionization (E$_{e^-}$= 50~eV) of methyl cyanide (CH$_3$CN) vapor, mass selected, and transferred to the cold ion trap. 
In the trap, the pure rotational transitions of \HCNH were measured employing a double-resonance vibrational-rotational spectroscopic scheme. 
Trap-based rotational techniques have been reviewed thoroughly \cite{asv21d},
and the particular scheme applied here was recently developed \citep{scm22a,asv23},
and already applied to molecular ions of astrophysical interest~\citep{silva23, gupta23}. 

\begin{figure} [h]
    \centering
    \includegraphics[width=1.0\linewidth]{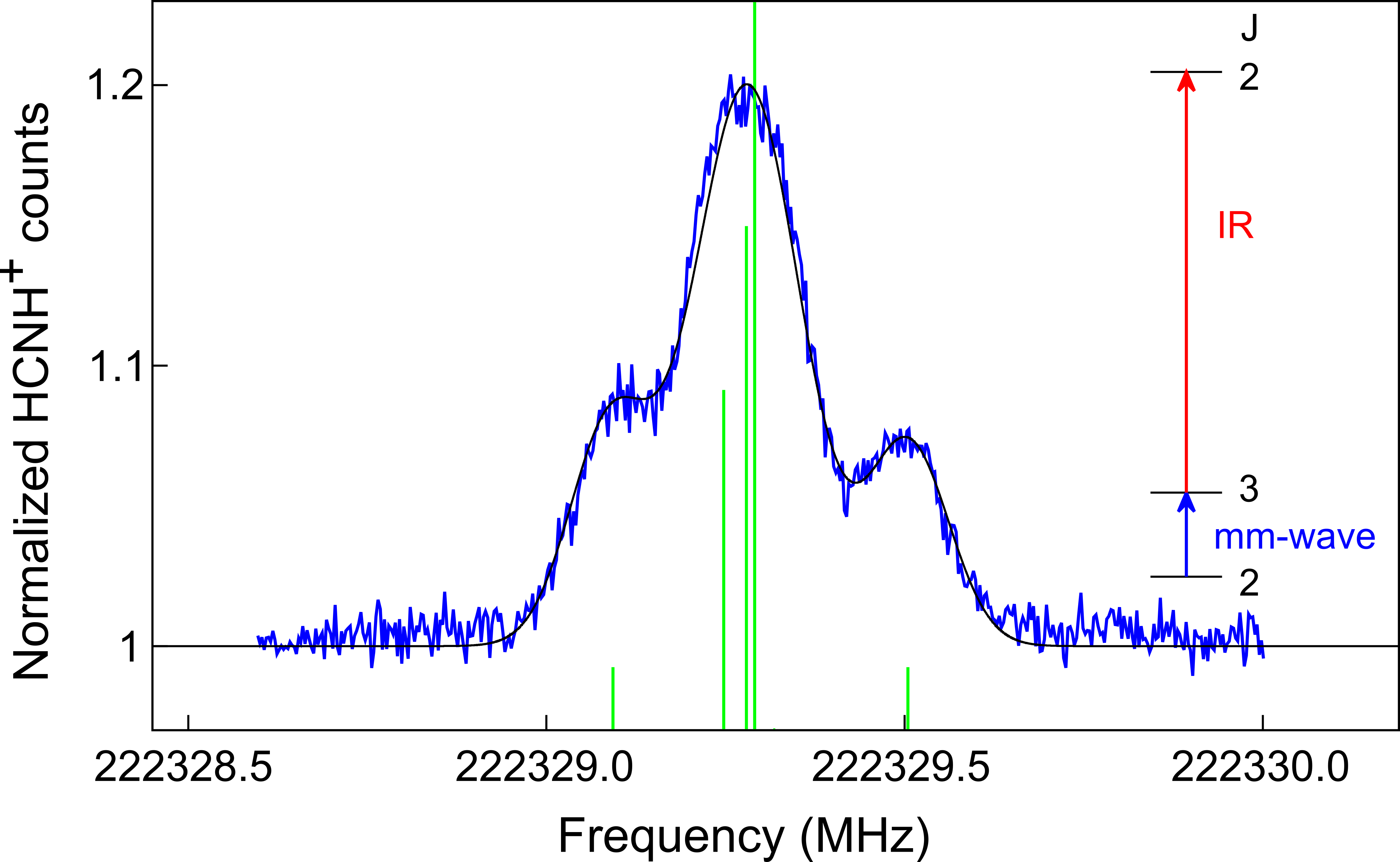}
    \vspace{-1.5em}
    \caption{Rotational transition (J= 3 $\leftarrow$ 2) of \HCNH showing partially resolved hyperfine structure recorded using double-resonance spectroscopy in a 4~K cryogenic ion trap. In this measurement, the IR laser was kept fixed on resonance with the $P$(3) rovibrational transition 
    (3180.401~cm$^{-1}$, Altman et al.\cite{alt84}) within the fundamental $\nu_2$ C-H stretch band. 
    The green sticks represent the simulated $^{14}$N hyperfine structure based on 
    the spectroscopic constants given in Table~\ref{param}. The shown 3-component Gaussian fit (black curve)
    yields a kinetic temperature of the ions of less than  20~K.
    Figures of other transitions can be found in the supplementary material file.} 
    
    \label{fig:enter-label}
\end{figure} 

An example of a rotational transition ($J= 3 \leftarrow 2$) recorded for \HCNH displaying partially resolved hyperfine splittings 
is shown in Fig.~1. While recording the rotational line, 
the wavenumber of the IR beam (red arrow in Fig.~1) is kept fixed on resonance with a 
rovibrational transition starting from a specific rotational level in the ground vibrational state. 
Then, millimeterwave radiation (blue arrow in Fig.~1) is used to excite a pure rotational 
transition starting or ending in the rotational quantum state probed by the IR laser, thus, increasing or decreasing the  signal counts.
For the IR excitation, selected rovibrational transitions within the fundamental $\nu_2$ 
C-H stretch band were used and they were readily identified based on the previous report by Altman et al.\cite{alt84}. 
The transition in Fig.~1  as well as all other rotational transitions were recorded in several individual measurements, 
in which the millimeterwave frequency was scanned back-and-forth in a given frequency window in constant steps. The step size was fixed
in each measurement and was typically 3-5~kHz. 
Care has been taken to lower the mm-wave power as much as possible to minimize power broadening effects.
The baseline in Fig.~1 was normalized following a frequency-switching procedure, 
where the \HCNH ion counts monitored in the frequency window of interest are divided by the \HCNH 
counts at an off-resonance frequency position. Thus, the baseline in the spectrum of Fig.~1 is close to unity.

Transition frequencies were determined by adjusting the parameters of an appropriate line function, 
typically a three-component Gaussian, in a least-squares procedure. 
In total, we measured six rotational transitions, 
the first four exhibiting resolved or partially resolved hyperfine structure. 
Their frequencies and uncertainties given in Table~\ref{rotlines}  are obtained from weighted averaging
of all available measurements. To obtain the accurate spectroscopic parameters 
reported in Table~\ref{param}, a global fit of our observed lines and 
those at higher frequencies previously measured by Amano et al.\cite{ama06} (also shown in Table~\ref{rotlines}) was carried out using a standard 
linear top Hamiltonian with a single quadrupolar nucleus as implemented in Western's PGOPHER program \citep{wes17}. We also performed a similar fit using Pickett's SPFIT/SPCAT program suite \cite{pic91} and the obtained values for the spectroscopic parameters match well with those from PGOPHER in Table \ref{param} within the error bars. The details of the SPFIT fit along with spectral predictions from SPCAT are provided as supplementary material. 
The spectroscopic parameters in Table~\ref{param} are considerably refined in this work and will be certainly useful for future astronomical observations. 
In particular, the $eQq$ value is now improved and based on a terrestrial measurement. 
Also, the  nuclear spin-rotation interaction constant $C_I$ is determined for the first time.

\begin{table} [h!]
\begin{center}
\caption{Ground state rotational transition frequencies of \HCNH (in MHz) and fit residuals  $o-c$ (in kHz).\label{rotlines}}
\begin{tabular}{ccr@{}lr@{}lr@{}l}
\hline
$J^\prime \leftarrow J^{\prime \prime}$ & $F^\prime \leftarrow F^{\prime \prime}$ & \multicolumn{2}{c}{Frequency $^a$} & \multicolumn{2}{c}{$o-c$}  \\
\hline
1 $\leftarrow$ 0 & 1 $\leftarrow$ 1   & 74111&.165(5)  &  $-$0.&3 \\
                 & 2 $\leftarrow$ 1   & 74111&.333(5)       &   $-$2.&2 \\
                 & 0 $\leftarrow$ 1   & 74111&.558(5)       &   0.&5       \\
\hline
2 $\leftarrow$ 1 & 2-2,1-0           & 148221&.284(15)     &  $-$13.&7  \\
                 & 3-2,2-1           & 148221&.462(5)      &     2.&0 \\
                 & 1 $\leftarrow$ 1  & 148221&.696(15)     &  $-$7.&4 \\
\hline
3 $\leftarrow$ 2 & 3$\leftarrow$ 3   & 222329&.092(15)      &    $-$1.&1 \\
                 & 2-1,3-2,4-3       & 222329&.279(5)       &     1.&9 \\
                 & 2 $\leftarrow$ 2  & 222329&.500(15)      &  $-$1.&2\\
\hline
4 $\leftarrow$ 3 & 4 $\leftarrow$ 4  & 296433&.445(15)      &     7.&1 \\
                 & 3-2,4-3,5-4       & 296433&.637(5)       &     1.&1 \\
                 & 3 $\leftarrow$ 3  & 296433&.842(15)      &     0.&5 \\
\hline
5 $\leftarrow$ 4 &                   & 370533&.362(5)          &  $-$2.&1 \\
6 $\leftarrow$ 5 &                   & 444627&.302(10)         &  $-$4.&9 \\
7 $\leftarrow$ 6 &                   & 518714&.331(25) $^b$    &    20.&5 \\
8 $\leftarrow$ 7 &                   & 592793&.222(10) $^b$    &     0.&8 \\
9 $\leftarrow$ 8 &                   & 666862&.895(25) $^b$    &    7.&3 \\
10 $\leftarrow$ 9 &                  & 740922&.154(25) $^b$    &  $-$6.&0 \\
\hline
\end{tabular}
\end{center}
 \raggedright 
\footnotesize{
$^a$ Former measurements from Refs~\cite{aka98,bog85} are not shown in this  Table\\
$^b$ From Amano et al. \cite{ama06}
}

\end{table}

\begin{table}
\begin{center}
\caption{\label{param} Spectroscopic parameters of ground state \HCNH, 
obtained by fitting the data given in Table~\ref{rotlines} with the program PGOPHER \cite{wes17}. 
All values are in MHz.}
\begin{tabular}{lr@{}lr@{}lr@{}l}
\hline
        Parameter $^a$ & \multicolumn{2}{c}{This work} & \multicolumn{2}{c}{Amano et al.\cite{ama06}} & \multicolumn{2}{c}{Ziurys et al.\cite{ziu92}}\\
\hline                    
  $B_0$              & 37055&.7482(3)   &    37055&.7518(12)  &    37055&.76(5)     \\
  $D_0 \times 10^3$  & 48&.248(9)        &      48&.234(107)   &    48&.4(11)   \\
  $H_0 \times 10^6$  &  0&.31(6)       &                      \\
  $eQq$($^{14}$N)    &  -0&.530(4)       &        &             &      -0&.49(7)\\
  $C_I$($^{14}$N)    &   0&.0053(8)      &   \\ 
  RMS                &     0&.0068            &    0&.035     &        0&.061 \\
\hline
\end{tabular}
\end{center}
 \raggedright 
\footnotesize{$^a$ Rotational constant ($B_0$), quartic ($D_0$) and sextic ($H_0$) centrifugal distortion constants, 
quadrupole coupling constant ($eQq$), and spin-rotation interaction~$C_I$\\}
\end{table}

\section*{Supplementary Material}
The PGOPHER and  SPFIT/SPCAT fit files are available as supplementary material, as well as Figures 
of the four lowest rotational lines.

\section*{Acknowledgments}
\vspace{-1em}
This work has been supported by an ERC advanced grant (MissIons: 101020583) 
as well as by the Deutsche Forschungsgemeinschaft (DFG) via 
Collaborative Research Center 1601 (project ID 500700252, sub-project C4) 
and "Schmid 514067452". 
W.G.D.P.S. thanks the Alexander von Humboldt Foundation for support through a postdoctoral fellowship.

\section*{Data Availability Statement}
\vspace{-1em}
Data available on request from the authors

\section*{References}
\vspace{-1em}
%

\end{document}